\documentclass[referee]{aa}

\usepackage{graphicx}

\usepackage{txfonts}

\begin{document}
\title{Are rotation curves in NGC 6946 and the Milky Way magnetically supported?}

\author{E. Battaner\inst{} \and E. Florido\inst{}}

\offprints{E. Battaner}

\institute{Departamento de F\'{\i}sica Te\'orica y del Cosmos. Universidad de Granada. Spain.\\
\email{battaner@ugr.es}
\email{estrella@ugr.es}}

\date{}

\abstract{ Following the model of magnetically supported rotation of spiral galaxies, the inner disk rotation is dominated by gravity but magnetism is not negligible at radii where the rotation curve becomes flat, and indeed becomes dominant at very large radii. Values of the order of 1 $\mu$G, or even less, produce a centripetal force when the absolute value of the slope of the curve [$B_\varphi$, R] (azimuthal field strength versus radius) is less than $R^{-1}$. The $R^{-1}$-profile is called the critical profile. From this hypothesis, the following is to be expected: at large radii, a ``subcritical'' profile (slope flatter than $R^{-1}$); at still larger radii a $B_\varphi$-profile becoming asymptotically critical as the density becomes asymptotically vanishing. Recent observations of magnetic fields in NGC 6946 and the Milky Way are in very good agreement with these predictions. This magnetic alternative requires neither galactic dark matter (DM) nor modification of fundamental laws of physics, but it is not in conflict with these hypotheses, especially with the existence of cosmological cold dark matter (CDM).
}
\titlerunning{Magnetic fields \& rotation}

\maketitle

\section{Introduction}

The standard interpretation of the fast rotation and the flat rotation  curves of spiral galaxies is based on the existence of massive and extended cold dark matter. This hypothesis is widely accepted, being itself a consequence of the so called CDM  paradigm. This paradigm provides a cosmological scenario in which a coherent interpretation of a large variety of phenomena is achieved, in particular concerning galaxy formation.

However, alternative interpretations of the rotation curve should be discussed, unless contradictions with observations require their withdrawal. Among such interpretations, Modified Newtonian Dynamics (MOND) is at present \ a subject of considerable debate. 

As a third possibility, the hypothesis of magnetically induced rotation  represents the simplest explanation, based on pure classical magnetohydrodynamics (MHD). Moreover, recent observations seem to  give strong support for this scenario. It is precisely the aim of this paper to draw the attention to the agreement between recent measurements of magnetic fields in NGC 6946 and the Milky Way and the general  predictions of the magnetic hypothesis.

After the pioneering work of Peratt (1986) and Nelson (1988), this model has been considered in a series of papers. In a first toy model, Battaner et al. (1992) showed that magnetic fields can provide a centripetal force. This first model proposed strengths of the order of 6 $\mu$G in the outer parts. This is too high for a galaxy like the Milky Way. It is, however, not too high for a galaxy like NGC 6946, with 10 $\mu$G at 10 kpc; this is the total field, and so the regular field could be about 5 $\mu$G. Among other works, let us highlight two important papers presenting arguments against our exploratory model: Cuddeford \& Binney (1993) considered that large magnetic field strengths would produce unacceptable flaring in the disk, while in the opinion of Jokipii \& Levy (1993) the Virial theorem implies a net centrifugal action. Both difficulties were surmounted in a more detailed two-dimensional model including vertical escape (Battaner \& Florido 1995; thereafter BF95) induced by magnetic fields in the disk. A mass loss rate of the order of 0.1 M$_\odot$yr$^{-1}$ prevented the excessive flaring, and the net action of magnetic fields was centrifugal, albeit centripetal in the radial direction. On the other hand, the magnetic strengths required were much lower, even less than 1 $\mu$G, a value perfectly acceptable in the outer regions of our Galaxy, even if our model was developed for any typical unspecified galaxy.  

A force cannot be ``a priori'' ignored unless it is at least one order of magnitude lower than the dominant force. In previous works we have shown that magnetic forces are not negligible at all in the dynamics of a galaxy, and indeed become dominant at large radii. This conclusion was also recently reached by Kutschera and Jalocha (2004). Further arguments in favour of the non-negligible influence of magnetic fields are given by Battaner \& Florido (2000). In smaller spirals, with a slower rotation and requiring more DM in the standard scenario, magnetic fields are important at all radii, as shown by Battaner et al. (2001) for DDO154. Attempting to deal with rotation curves of spirals, without taking into account magnetic fields, may be completely unrealistic. These order-of-magnitude arguments will not be repeated here, but rather we will focus on the agreement with recent observations.

Even if the model in BF95 calculates magnetic field strengths, densities, radial and vertical winds and escape, in reality the dynamics of the interstellar medium (ISM) is very complicated, mainly due to the combined actions of turbulence, finite electrical conductivity, supernova (SN) explosions and other non-negligible ingredients. Therefore, the magnetic model to explain rotation curves (like any other model with this objective) is still at an embryonic stage. However, idealistic models often provide light and complicated small-scale transient phenomena are diluted  when we consider large length and time scales, mainly in the most external regions. In any case, other models, such as DM and MOND, also require currently the same level of simplicity.

Small scale dynamics (say, less than 5-10 kpc) may be important in the bulk motions of the galaxy. Some related phenomena cannot be neglected and these facts should be considered in future realistic models.

Supernova explosions are the main source of turbulence. A single SN ejects about $10^{51}$ erg, which is important given the high rate of SN explosions and that the kinetic energy of galactic rotation is of the order of $10^{59}$ erg. 3D simulations of the galactic turbulence by Korpi et al. (1999) using the code of Brandenburg et al. (1995) show that about 9\% of  the SN energy is transferred to the kinetic energy of the ISM. These authors reproduce the way in which two main phases are developed, the warm and the hot phases, together with some particular features of  the cold gas. It is clear that magnetic fields, partially frozen-in, in a complicated multiphase medium, in which different phases present large supersonic and superalfvenic velocities with respect to one another, are an important dynamical phenomenon, even if difficult to deal with.

Turbulence has several effects rendering magnetic fields a complicated matter. It is the fundamental mechanism for dynamos, without which turbulent magnetic diffusion would dissipate large scale fields. Turbulent magnetic diffusion may, in turn, produce an inward and/or outward transport in the vertical direction, enabling an interchange of fields between the ISM and the intracluster medium. Magnetic reconnection is also a very important effect, as demonstrated by Lazarian and Vishniac (1999), not only in establishing the large scale field but also as a heating source.

Turbulence must be specially important in galaxies with very active star formation and a high SN rate. In the galaxy NGC 6946, the magnetic field strength has been measured very precisely, and so it is very appropriate for comparison with theoretical outputs. However, it has a very large SN rate, rendering the interpretation of its dynamics less easy.

A realistic model of the ISM dynamics, therefore, must take into account all these effects. For smaller scales, the 3D simulations of Korpi et al. (1999) deal with this complicated picture. In the model the volume considered is only (0.5x0.5x2)kpc$^3$ (2 kpc in the vertical direction). The authors also considered relatively short integration times, less than 100 Myr, as otherwise, the escape becomes too large. However, recent MHD simulations on larger scales (Avillez \& Breitschwerdt, 2005) do not show the deficiencies of small computational volume and small integration time.

We expect (and our  simplification could be useful) that  at larger space and time scales, the whole picture must become much simpler. Magnetic field lines become nearly azimuthal, even if the mean circular line is very corrugated and twisted at smaller scales.

In any case, it is an observational fact that spiral galaxies have large scale magnetic fields with a predominant azimuthal component, $B_\varphi$, which are so high that they have to be taken into account when considering the rotation curve of spiral galaxies. Recent measurements by Mitra et al. (2003) and Brown et al. (2003) confirm that also in the Milky Way the large scale magnetic field distribution is as regular as in other spiral galaxies.

It is important to realise that in this paper we intend neither to develop a new model, nor to take our previous model to be compared with observations. We only need to consider that the net magnetic force is
\begin{equation}
   F_{mag}= -{1 \over {8\pi R^2}}{{\partial \left( R^2 B_\varphi^2 \right)} \over {\partial R}} 
\end{equation}

 Let us briefly derive this expression in order to highlight the assumptions required. Under the conditions leading to cosmic MHD, the fluid motion equation can be written as
\begin{equation}
  \rho {{\partial \vec{v}_o} \over {\partial t}} +
  \rho \vec{v}_o \cdot \nabla \vec{v}_o + \nabla P =
  n \vec{F} + {1 \over {4 \pi}} \vec{B} \cdot \nabla \vec{B} -
  \nabla {{B^2} \over {8 \pi}}
\end{equation}
 where $\rho$ is the density, $\vec{v}_o$ the velocity, $P$ the pressure, $n$ the number density, $\vec{F}$ any non-magnetic force and $\vec{B}$ the field strength. Therefore, we are assuming (as usual) the standard set of MHD conditions. The most problematic assumption is that of infinite conductivity, which implies vanishing diffusion effects, or, equivalently, a perfect coupling between field and fluid. This assumption requires careful discussion, given the highly complicated behaviour of the ISM, but can be accepted in a large-scale ideal description.

This coupling is usually assumed, but is only in part theoretically justified. Some recent observations provide evidence of such a coupling, at least in the warm medium (Mitra et al. 2003), having the warm and the cold medium the same distribution at large scales (Walsh et al. 2002).

 In cylindrical coordinates the magnetic force, from the above equation, is
\begin{equation}
  \vec{F}_{mag}= {1 \over {4\pi}} \vec{B} \cdot \nabla \vec{B} -
                 \nabla {{B^2} \over {8 \pi}} =
                 {1 \over {4\pi}} \left( B_R, B_\varphi, B_z \right)
     \left[ \begin{array}{lll}
        {{\partial B_R} \over {\partial R}} & {{\partial B_\varphi} \over {\partial R}} & {{\partial B_z} \over {\partial R}} \\
            &        &                     \\
       {1 \over R}{{\partial B_\varphi} \over {\partial B_\varphi}} - {B_\varphi \over R} & {1 \over R}{{\partial B_\varphi} \over {\partial B_\varphi}} + {B_R \over R} & {1 \over R} {{\partial B_z} \over {\partial B_\varphi}} \\
            &        &                     \\
{{\partial B_R} \over {\partial z}} & {{\partial B_\varphi} \over {\partial z}} & {{\partial B_z} \over {\partial z}}
       \end{array} \right]-
     {1 \over {8 \pi}} \left[ {{\partial B^2} \over {\partial R}}, {1 \over R} {{\partial B^2} \over {\partial \varphi}}, {{\partial B^2} \over {\partial z}} \right]
\end{equation}

 We further assume $B_R = B_z = 0$ and $\partial / \partial \varphi =0$, i.e. only the toroidal component exists and we have cylindrical symmetry (these assumptions are not necessary; they simply  allow us to focus on the fundamental properties).  We then have
\begin{equation}
  \vec{F}_{mag} = { 1 \over {4 \pi}} (0, B_\varphi, 0) \left( \begin{array}{ccc}
    0             & {{\partial B_\varphi}\over {\partial R}} &     0       \\
   -{{B_\varphi} \over R} &       0                         &     0       \\
    0             & {{\partial B_\varphi}\over {\partial z}} &     0  
   \end{array} \right) - {1 \over {8\pi}} 
   \left({{\partial B_\varphi^2} \over {\partial R}}, 0, 
         {{\partial B_\varphi^2} \over {\partial z}} \right)
\end{equation}

Therefore, the radial component of interest here is
\begin{equation}
  F_{mag} = - {1\over {4\pi}} \left( {{B_\varphi^2} \over R} + B_\varphi {{\partial B_\varphi} \over {\partial R}}\right)
\end{equation}
which is equivalent to eq. (1).

 This formula represents the actual radial magnetic force        
without invoking any particular model. With this simple formula in mind we are able to show that the  general scenario precisely matches what is observed. Our conclusions are, therefore, general, and not subject to the particular assumptions that any theoretical model must assume.  

\section{Agreement between theoretical predictions and recent measurements}

Following Beck (2004b), the total (regular plus random) value of B in our Galaxy is $\sim 4 \mu G$ at 15 kpc. For $B_{regular}/B_{total} > 0.6$ (Berkhuijsen 1971; Heiles 1996) a regular azimuthal field larger than 2$\mu$G is a conservative estimate. The value of $B_{regular}/B_{total}$ increases with radius and is larger than 0.6 in the outer parts. In the case of NGC6946, for which very precise measurements are available (Beck 2004a), magnetic energy densities are $10^{-11} erg cm^{-3}$ at 4 kpc (see fig. 1).

These strengths are of the order of magnitude needed for magnetic forces to become important and even dominant at very large radii. The radial scale length of synchrotron emission in nearby galaxies is much larger than that of other star formation indicators, and therefore magnetic fields must extend to very large radii (Beck 2004a).

As shown in Battaner et al. (1992) and in BF95, two magnetic forces have opposite directions: the force of the magnetic pressure gradient, which should be centrifugal, and the magnetic tension, which, in disk-shaped galaxies, is always centripetal. It is easily shown that the centripetal force predominates when the slope of the profile [B$_\varphi$,R] is lower than a ``critical slope''. In fact, the radial component of both magnetic forces per unit volume can be written, in axisymmetrical conditions,  as in equation 1.

 This critical slope corresponds to $B_\varphi \propto 1/R$. ($B_\varphi$ is the azimuthal component of $\vec{B}$). Let us call $B_\varphi \propto 1/R$ the ``critical profile'', for which the magnetic force vanishes.

In a typical spiral with a magnetically driven rotation, we can distinguish three regions: 

1) An internal one in which magnetism cannot compete with gravity; 

2) An intermediate one with a ``sub-critical'' $B_\varphi$-profile, i.e. where the absolute value of the slope of the curve $[B_\varphi, R]$ is lower than the 1/R-profile. (More precisely when $-{1 \over B}{{dB} \over {dR}} < {1 \over R}$). 

3) An external one, in which the $B_\varphi$ profile becomes asymptotically critical. This is because the density asymptotically decreases to zero (nearly exponential), i.e. if the density $\rho (R \rightarrow \infty) \rightarrow 0$, we should expect $F_{magnetic} (R\rightarrow \infty) \rightarrow 0$, too. Therefore, in the outermost disk we expect a critical profile $B_\varphi \propto 1/R$.

The theoretical profile for an unspecified spiral, calculated by BF95, gives also a critical profile at large radii. Any magnetic model would predict a critical profile (1/R) at very large radii.

Fig. 1 plots the magnetic energy density as deduced from observations in NGC 6946 (Beck 2004b) together with the critical $R^{-1}$-profile passing through the point at 4 kpc. It is evident that the $B_\varphi$-profile is sub-critical for $R \ge 3$ kpc, i.e. the net magnetic force is inwards. Following Beck, the field energy density has a radial scale length that is much larger than the neutral density scale length of only $\sim$ 3 kpc.

The slow decrease of $B_\varphi$ with $R$ is able to produce a very large centripetal force. At 8 kpc this force can be easily calculated to be the same as  that produced by a central point mass of $5 \times 10^{9} M_\odot$, which is comparable to the visible mass for R $<$ 8 kpc (Battaner et al. 2005). The third region with critical slope has not been observationally reached yet.

 Equipartition between fields and cosmic rays is probably still valid in the outer regions. Even if there is no star formation to produce cosmic rays there, the cosmic ray energy density cannot be neglected at a distance from the sources, as the particles can diffuse considerable distances and are specially abundant above the galactic disc. The energy density of cosmic rays is dominated by the nucleon component for which energy losses by synchrotron or inverse Compton are negligible and  which can therefore be found at large distances from its birth place. However, when calculating the synchrotron emission by electrons, these energy losses have to be taken into account, with 1 kpc being a typical distance for an electron to become thermal.

The turbulent energy density is one order of magnitude smaller than the magnetic energy density. Equipartition between the two energy densities does not hold; thus, the larger the radius the greater the difference. This is in disagreement with most models assuming a turbulent generation of the galactic field. The hypothesis of turbulent magnetic diffusion providing equilibrium between the z-components of both galactic and intergalactic fields, followed by an $\Omega$-effect, could be a possible explanation. This simple mechanism provides a critical profile, as shown by Battaner \& Florido (2000). See also Breitschwerdt et al. (2002), and arguments against the inward flux by magnetic diffusion in Ferri$\grave{e}$re (2004).

Fig. 2 plots the total field in the Milky Way for R$<$17 kpc. The observational profile is taken from the review by Beck (2004a). It is based on measurements by Beuermann et al. (1985) assuming equipartition between magnetic field and cosmic ray energy densities. To obtain regular fields, these values should be multiplied by a factor $\sim 0.6$. Together with the observational values we also plot the critical $R^{-1}-$profile passing through the point at 10 kpc. 

It is evident that the actual profile very closely matches the critical profile at large radii, as predicted by the magnetic alternative. Before the critical slope is reached, it is apparent that the profile is ``sub-critical'', in agreement with our theoretical expectations.
In the Milky Way, therefore, the gravitation dominated region would be identified with $R \lesssim 7$ kpc, the subcritical region with centripetal magnetic force for $7 \lesssim R \lesssim 10$ kpc and the third region with critical $B$ profile for $R \gtrsim 10$ kpc. That is clearly appreciated in Fig. 2.
Magnetic forces become larger at larger radii because the radial scale length of the field energy density is 8 kpc while the density scale length is only 3 kpc (Beck 2004a).

 Observe that in Fig. 1, for the case of NGC 6946, we plot magnetic energy densities ($B^2/8\pi$) while in Fig. 2, we plot the magnetic field strength ($B$) in the case of the Milky Way. This difference is because we want to keep the cited observational results in the original form in which they were presented.

Large magnetic strengths in the outer disk of M31 were observed by Han et al. (1998), who found that regular fields at 25 kpc were similar to those observed in the inner disk. This means not only that strengths in the external region of M31 are large, but also that the slope is very low, i.e. they are perfectly able to produce an important centripetal force.

Therefore, both observations of the field strength and the low slope of the curve [$B_\varphi$,R], give support to the magnetic model of the rotation curve of spirals.

\section{Discussion}

The magnetic hypothesis to explain the rotation curve is not in  fundamental disagreement with current models on cosmology and galaxy formation. Rejecting galactic DM does not imply rejecting DM in the Universe. Galactic DM, even if very important in the dynamics of a galaxy, would only provide $\Omega \sim$ 0.024  (Fukugita \& Peebles 2004), less than one order of magnitude than the currently assumed value of DM in the Universe. Therefore, DM in the Universe and DM in a galaxy are relatively unconnected problems. Probably the PLANCK mission  will provide new insight into this problem.

The pure magnetic model, without galactic DM, is, however, in clear contradiction with current simulations of galaxy formation, as they predict large amounts of galactic DM distributed following the so called NFW profiles (Navarro et al. 1996, 1997). It is more compatible with early analytical CDM models and the scenario conceived by White \& Rees (1978) in which only pre-galactic baryonic concentrations would survive the violent merger events, DM being redistributed and erasing most of the DM substructure. 

Having remarked upon this disagreement, however, the inclusion of magnetic fields in the rotation analysis could solve some problems in CDM (or $\Lambda$CDM) models. These models do not satisfactorily reproduce the observational rotation curves (Navarro \& Steinmetz 2000; Hayashi \& Navarro 2002; Abadi et al. 2003). Neither do they satisfactorily explain the Tully-Fisher relation (which, in turn, is not the intention of the present paper). The absence of a correlation between the asymptotic velocity and the orbital velocities of satellites would be simply explained by the magnetic alternative, the former being the result of gravity plus magnetism, the latter, of gravity alone. Other advantages of introducing magnetism to study rotation properties of galaxies are given by Battaner \& Florido (2000).

An additional argument in favour of the magnetic model, considered by Kutschera and Jalocha (2004), is the fact that elliptical galaxies, in which magnetic fields are clearly negligible, show  Keplerian rotation curves. Romanovsky et al. (2003) showed that the rotation curves of intermediate luminosity ellipticals do not require the presence of DM.

\section{Conclusions}

We are aware that a complete description of galactic magnetic fields and their dynamical influence must take into account dissipative effects, fountains, infall and galactic winds, and cosmic rays, among other effects, which renders the use of ideal MHD questionable. An effective coupling between fields and ISM is necessary for the model of magnetically driven rotation curves to work. This coupling is currently assumed in cosmic MHD but an observational support is necessary. Even so, in this case ideal models provide a reasonable interpretation of the observed magnetic fields.

Recent data about regular magnetic fields in spiral galaxies have been presented by Beck (2004b) in a recent review that clearly confirms what is to be expected in the magnetic scenario for rotation curves: i.e. external magnetic fields of about 1 $\mu$G,  with slopes of its azimuthal component vs. radius being flatter or very slightly flatter (observationally inappreciable) than the $B_\varphi \propto R^{-1}$-critical slope.

The magnetic alternative remains a serious, competitive theory. It requires neither the existence of DM nor the modification of classical laws (including General Relativity). It is based on MHD, a relatively recent chapter of Astrophysics, but one that has roots in classical electromagnetism. It is to be emphasised that this alternative does not imply that DM does not exist, but just that it should be located outside galaxies, probably (and mainly) in clusters. Therefore, the magnetic alternative is not in  fundamental conflict with the broadly accepted cosmological scenario with CDM and dark energy. Rather, the inclusion of magnetic effects, which is in any case necessary, could help to theoretically reproduce some unexplained, well known facts, for example, the rotation curve. Gravity alone does not explain the rotation curve very well, simply because magnetic fields cannot be ignored.

It is true that simulations of galaxy formation predict big halos around galactic baryonic concentrations and the magnetic model does not require them. Galaxies with a rotation dynamics dominated by both dark halos and magnetic fields cannot be rejected. The alternative magnetic model provides an interpretation of all observations alone, without hybrid solutions, although hybrid models cannot be disregarded.

 The dynamic role of galactic magnetic fields is a matter that can no longer be ignored, neither at the small nor at the large scale.

\begin{acknowledgements}

We are very grateful for the information, valuable comments and discussions provided by Dr. R. Beck and Prof. R. Wielebinski. This paper has very much benefited from their help.

We would like to thank the anonymous referee for constructive comments and suggestions. We would also like to thank Ute Lisenfeld for the correction of the manuscript.

This paper has been supported by the ``Plan Andaluz de Investigaci\'on'' (FQM-108) and by the ``Secretar\'{\i}a de Estado de Pol\'{\i}tica Cient\'{\i}fica y Tecnol\'ogica'' (AYA2000-1574, AYA2004-08251-C02-02, ESP2004-06870-C02-02).
\end{acknowledgements}

\clearpage

\begin{figure}
\includegraphics{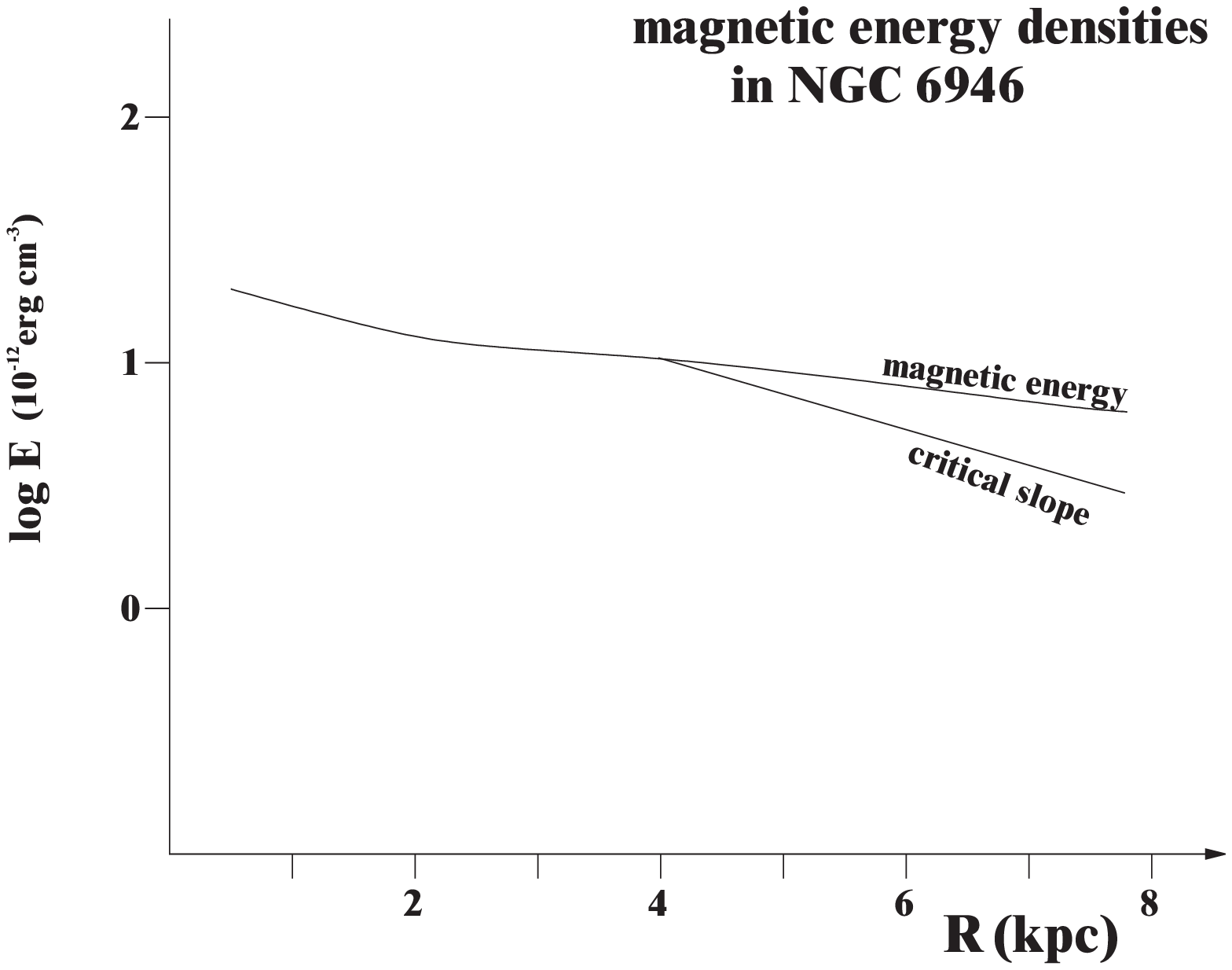}
\caption{Observational magnetic energy density for NGC 6946 from Beck (2003). We add a curve showing the critical slope passing through the observational point at R= 4 kpc. It is appreciated that for R$\ge$4 kpc the true slope is subcritical producing a large inward force on the gas.}
\end{figure}

\begin{figure}
\includegraphics{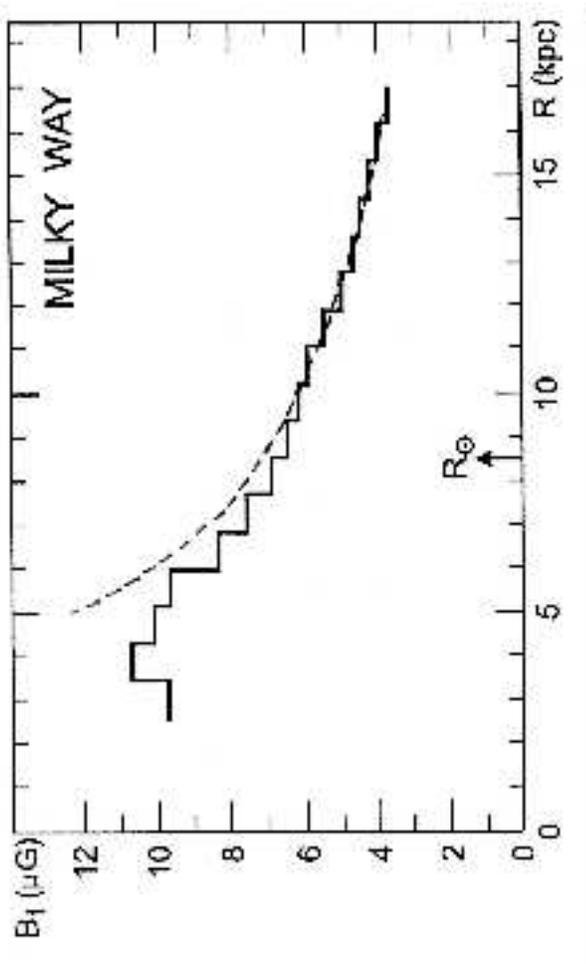}
\caption{Observational magnetic field strength in the Milky Way, from Beck (2003) (solid curve). The dotted curve represents the critical slope passing through the observational point at R=10 kpc. It can be appreciated that the real profile is critical for R$\ge$ 10kpc and subcritical between about 7 and 10 kpc.}
\end{figure}

\end{document}